# Auger decay in krypton induced by attosecond pulse trains and twin pulses

**Christian Buth**[1], **Kenneth J. Schafer**[2]

Department of Physics and Astronomy, Louisiana State University, Baton Rouge, Louisiana 70803, USA

**Synopsis:** Using attoscience, we study the electron correlations responsible for Auger decay in krypton atoms. The Auger decay is induced by a pulse train or a twin pulse composed of subpulses of attosecond duration. During the Auger decay an optical dressing laser may be present. Interference effects between the ejected Auger electron wave packets are predicted.

Attosecond science aims at studying and controlling the motion of electrons on their natural time scale. The investigation of electronic motion has so far been restricted mostly to one-electron processes. One of the first applications of attoscience was the measurement of the Auger decay time of 3d vacancies in krypton atoms—a well-known datum from frequency-domain spectroscopy—with a single attosecond pulse in the presence of an optical streaking laser.

In the present work, we examine the Auger electron spectrum of krypton induced by an arbitrarily shaped XUV light pulse with and without an optical dressing laser. Specifically, we consider Auger decay after

1. a train of attosecond pulses with a set duration between the pulses and
2. an attosecond twin pulse with varying time delay between the subpulses: an archetypical interference experiment.

The first point refers to a proposed experiment, which will repeat the measurement of the Auger decay lifetime of krypton 3d vacancies with pulse trains [1]. A train produces a stronger signal than a single pulse and enhances the signal to noise ratio in an experiment. Further a train of pulses leads to electron wave packet interference of the wave packets launched by different pulses. An additional optical streaking laser facilitates the control and measurement of the resulting signal. The second point proposes an experiment with a simpler setting compared with a pulse train. Additional flexibility is added by varying the time delay between the two attosecond pulses. Predictions of the second setting are displayed in Fig. 1. The upper panel shows the attosecond twin pulse and the exponential decay of the Auger population from a rate equation model in the time domain. After a pulse, the population of inner-shell ionized atoms decays exponentially. The lower panel shows the Auger electron spectrum from our quantum dynamical model [2,3], if the photoelectron energy is observed in coincidence with the Auger electrons with the stated energy resolution. We predict interference between the two ejected Auger electron wave packets.

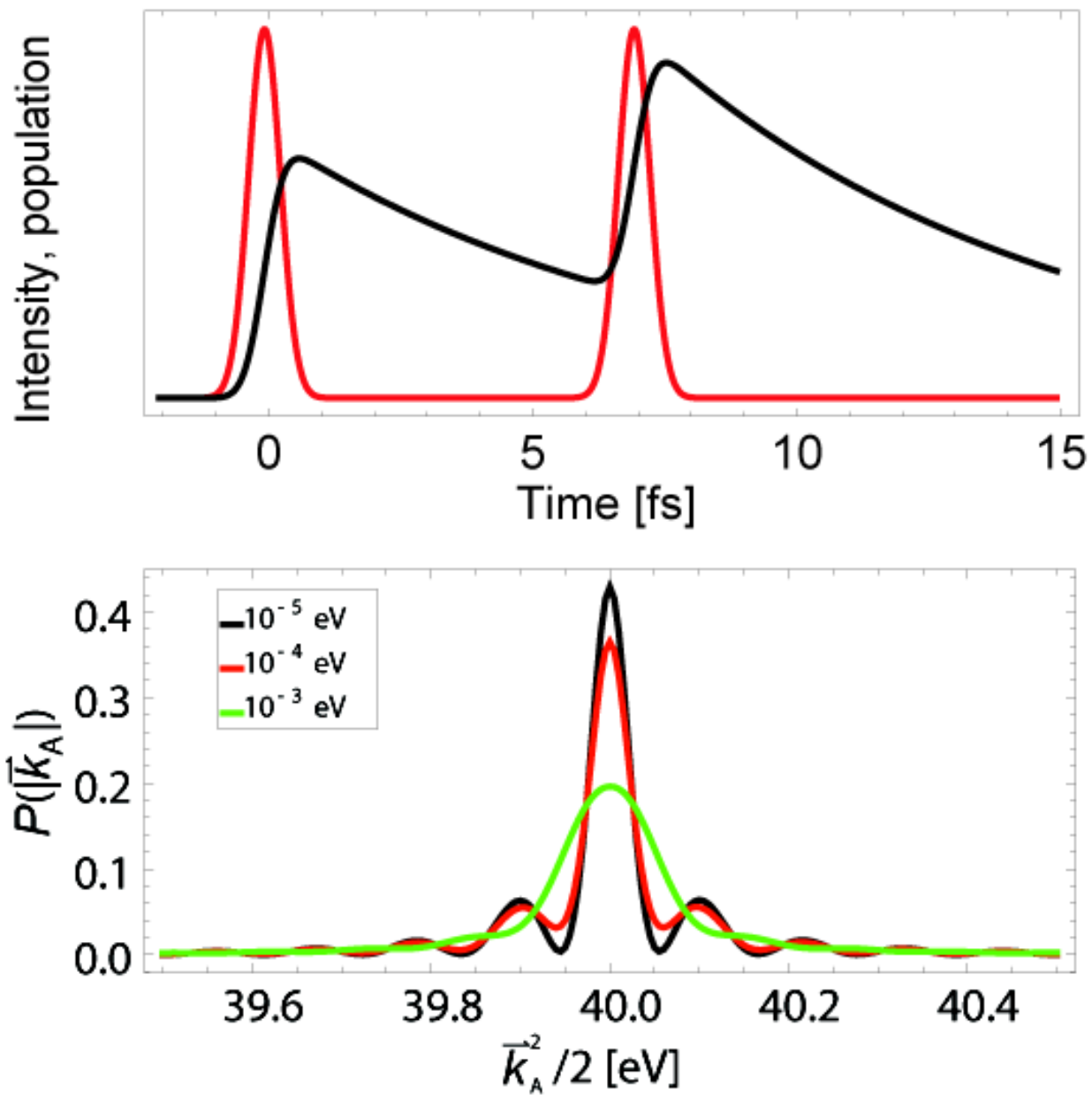

**Fig. 1.** (Color) Upper panel: intensity of an XUV twin pulse (red) and the krypton 3*d* hole population (black). Lower panel: probability density of Auger electrons with momentum $\vec{k}_A$ in the opposite direction of the linear XUV polarization vector for three energy resolutions of the photoelectron measurement.

In conclusion, our proposed experiments are very fundamental; they represent a crucial test for attoscience: to what degree can we measure and control an electronic process in the time domain?

## References

[1] J. Mauritsson, private communication (2009).
[2] C. Buth and K. J. Schafer, in preparation (2009).
[3] O. Smirnova, V. S. Yakovlev and A. Scrinzi, *Phys. Rev. Lett.* **91**, 253001 (2003).

[1] E-mail: christian.buth@web.de
[2] E-mail: schafer@phys.lsu.edu